\documentclass[11pt,twoside]{article}
\usepackage{./asp2010}

\resetcounters

\begin{document}

\title{The white dwarf cooling sequence of the Galactic bulge\footnotemark}

\author{Annalisa Calamida,$^1$ Kailash Sahu,$^1$ Jay Anderson,$^1$ Stefano Casertano,$^1$ 
Tom Brown,$^1$ Santi Cassisi,$^2$ Josh Sokol,$^1$
Howard Bond,$^1$  Harry Ferguson,$^1$  Mario Livio,$^1$ Maurizio Salaris,$^3$ and
Jeff Valenti$^1$
\affil{$^1$ Space Telescope Science Institute - AURA}
\affil{$^2$Osservatorio Astronomico di Teramo - INAF}
\affil{$^3$Astrophysics Research Institute, Liverpool John Moores University}
}

\titlefootnote{Based on observations made with the NASA/ESA \emph{Hubble Space Telescope}, obtained at STScI, which is operated by AURA, Inc. }

\begin{abstract}
We collected $F606W$- and $F814W$-band time-series data of the Sagittarius low-reddening window 
in the Galactic bulge with the Advanced Camera far Surveys mounted on the Hubble Space Telescope. 
We sampled the region approximately every two weeks for one year, with the principal aim to detect 
a hidden population of isolated black holes and neutron stars in the Galactic disk through astrometric microlensing.
We present preliminary results here based on a photometric catalog including 
$\approx 3 \times 10^5$ stars down to $F606W \approx$ 31 mag. 
Proper motions were also measured, with an accuracy of better than 
$\approx$ 0.5 mas/yr at $F606W \approx$ 28 mag in both coordinates. 
We were then able to separate disk and bulge stars and to obtain a clean bulge color-magnitude diagram.
Together with a dozen candidate extreme horizontal branch stars we were able to identify for the first time 
a clearly defined white dwarf (WD) cooling sequence in the bulge.
The comparison between theory and observations shows that a substantial fraction of the WDs ($\approx$ 40\%) 
is systematically redder than the canonical cooling tracks for CO-core DA WDs.
This evidence would suggest the presence of a significant number of He-core WDs in the bulge, formed
in close binaries, as has been found in some Galactic globular and open clusters. The presence of close 
binaries in the bulge population is further supported by the finding of a candidate dwarf nova in outburst 
and a few candidate cataclysmic variables in quiescence in the same field.
\end{abstract}

\section{Introduction}\label{intro}
The characterization of white dwarf (WD) properties, such as their luminosity and
temperature, is a valuable tool to understand the formation history of the different components of our Galaxy, 
since most stars end up their evolution as WDs \citep{hansliebert03}. 
Galactic-disk WDs have been extensively observed through imaging and characterized through spectroscopy \citep{eisenstein06, kepler07,koester09}. An updated catalog from the Sloan Digital Sky Survey (SDSS) Data Release 7 including 12,843 DA and 923 DB field WDs has recently been presented by \citet{kleinman13}. They find a mean mass of 
$\sim$ 0.6$M_{\odot}$ for DA and $\sim$ 0.68 $M_{\odot}$ for DB WDs. 
There also is a secondary peak in the disk WD mass distribution around 0.4 $M_{\odot}$ \citep{kepler07}, 
with $\approx$ 4\% of WDs being less massive than this value \citep{rebassa11}. These low-mass WDs 
probably have Helium cores and, given that their ages are less than a Hubble time, they
must result from close-binary interactions after a post-common envelope phase or from the merging of
two very low-mass He-core WDs \citep{han08}. The binary scenario is supported by the finding of numerous
low-mass WDs in binary systems in the field, with another WD as a companion, or a neutron star 
or a subdwarf B star \citep{maxted02,marsh95}. He-core WDs have also been observed in Galactic 
globular clusters, such as $\omega$ Cen \citep{monelli05, calamida08, bellini13}, NGC~6752 \citep{ferraro03}, and 
NGC~6397 \citep{strickler09}, and in the old
metal-rich open cluster NGC~6791 \citep{kalirai07}. He-core WDs in clusters show systematically 
redder colors compared to CO-core WDs. Moreover, since the cooling time scales of He-core WDs with mass
$\approx$ 0.4 $M_{\odot}$ are significantly longer compared to those of CO-core WDs, an excess of bright 
WDs should be observed compared to what predicted by DA CO-core models, as found in $\omega$ Cen \citep{calamida08}. 
At the same time, a population of extreme horizontal branch (EHB) stars is observed in most of these clusters, 
also in a metal-rich environment as NGC~6791 \citep{kalirai07, bedin08}.

Until now, a WD cooling sequence had never been observed in the bulge before, even though EHB stars 
have been identified in Baade's window by \citet{zoccali03}, and spectroscopically characterized as hot subdwarf stars by \citet{busso05}. The origin of EHB stars in the field and in clusters is still debated: while almost all of them seem to be binaries in the field \citep{maxted01}, they could follow different evolutionary channels in globular clusters, i.e.
they could be the progeny of $He$-enriched stars \citep{lee05}, or the aftermath of 
hot-helium flashers \citep{castellani93, dcruz96, cassisi03, castellani06}, or the merger of two WDs \citep{han08}. 
Similar scenarios are then proposed for the formation of He-core WDs in Galactic clusters.
Understanding the origin of EHB and WD stars in the bulge, and charactering their properties, is then fundamental 
to constrain the formation history of the bulge stellar populations and the Milky Way.

\section{Observations and photometry}
We observed the SWEEPS (Sagittarius Window Eclipsing Extrasolar Planet Search) field
(RA = 17:59:01, DEC = -29:12:00) in the Galactic bulge in 2004 and 2012 with the Hubble 
Space Telescope (HST) using the Advanced Camera's Wide-Field Channel 
(proposals GO-9750 and GO-12586, PI: Sahu). 
The 2004 observations were taken over the course of one week and were collected in the $F606W, F814W$ 
filters (for details about these observations and data reduction see \citealt{sahu06}). 
The new data were collected from October 2011 and October 2012, with a two week cadence, for a
total of 29 $F606W$- and 30 $F814W$-band images. 
The SWEEPS field covers an area of $\approx$ 3.3\arcmin $\times$ 3.3\arcmin;
it is a region of relatively low extinction in the bulge (E(B-V) $\lesssim$ 0.6 mag, \citealt{oosterhoff}).
Data from 2012 were reduced using an in-house software program that performed simultaneous
point-spread function (PSF) photometry on all 59 $F606W$ and $F814W$-band images (Anderson et al, in prep).
We adopted the 2004 photometric zero-point to calibrate the data to Vegamag, ending up with a catalog 
of $\approx$ 280,000 stars, down to $F606W \approx$ 31 mag. This is the deepest Color-Magnitude Diagram (CMD) 
ever obtained in the direction of the Galactic bulge. Our reduction software provided us with some quality parameters we adopt to select the photometry, such as the Root Mean Square (RMS) of the individual photometric and astrometric observations 
about the mean, the similarity of the object to the shape of the PSF ($q$) and
the degree of contamination of the object by neighboring stars ($o$).

\begin{figure}
\center
\includegraphics[width=12truecm,height=11truecm]{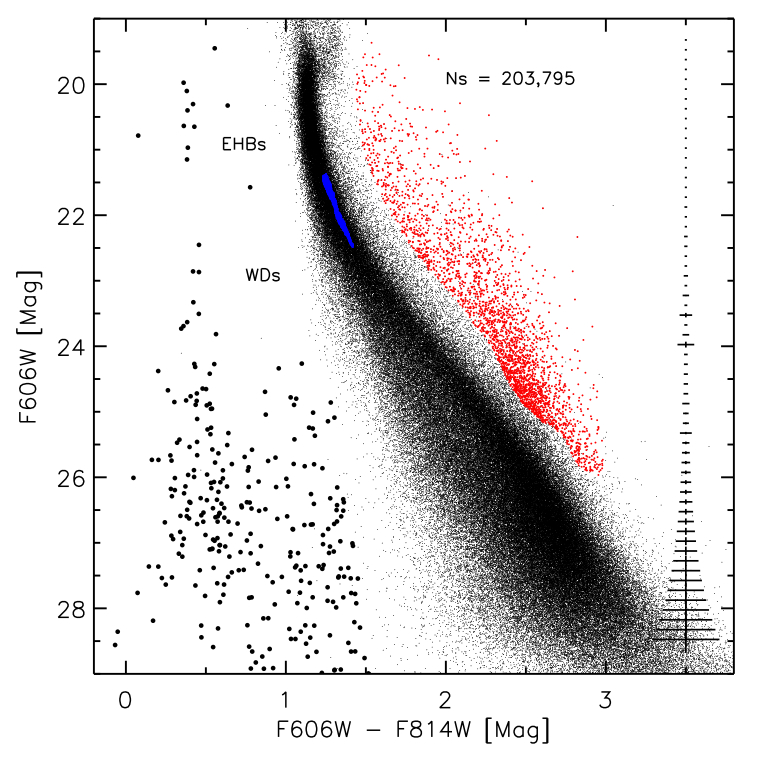} 
\caption{\label{fig1} $F606W,\ F606W - F814W$ CMD of selected stars in the SWEEPS field. The 
EHB and the WD stars are marked with larger filled dots, while candidate bulge and disk stars
are marked with blue and red dots, respectively.}
\end{figure}

Fig.~1 shows the $F606W, F606W - F814W$ CMD of a sample 
of $\sim$ 204,000 stars in the SWEEPS field (black dots). Stars were selected 
based on the quality of photometry:  within each 0.15-magnitude bin, we threw out the 20\% of 
the objects that had the most neighbor contamination (the $o$ parameter) and threw out the 
worst 10\% of the objects in terms of quality-of-fit ($q$). Stars brighter than $F606W \approx$ 19.5 mag are
saturated and we reach $F606W \approx$ 28.5 mag with a Signal to Noise ratio of $\approx$ 5 and
a completeness of $\approx$ 30\% (see sec. \ref{pm}).
The CMD shows some interesting features: the bulge main sequence (MS) presents a color spread much larger
than the spread given by photometric errors only; there is a group of stars systematically redder than the 
bulge MS that probably belongs to the (closer) disk population; a dozen candidate extreme horizontal branch (EHB) stars are clearly visible in the magnitude range  20 $\lesssim F606W \lesssim$ 21.5 mag and for $F606W - F814W \approx$ 0.3 mag;
a well-populated WD cooling sequence starting at $F606W \approx$ 22.5 mag and in the color range
0 $\lesssim F606W - F814W \lesssim$ 1.5 mag is visible. 
EHB stars were previously identified towards Baade's window (see sec. \ref{intro}), but this is the first time 
that a WD cooling sequence has been observed towards the bulge. It is worth noting that the sequence includes 
disk and bulge WDs.

\begin{figure}[htbp]
     \begin{minipage}[l]{0.48\textwidth}
      \centering
       \includegraphics[height=6.5cm,width=7cm]{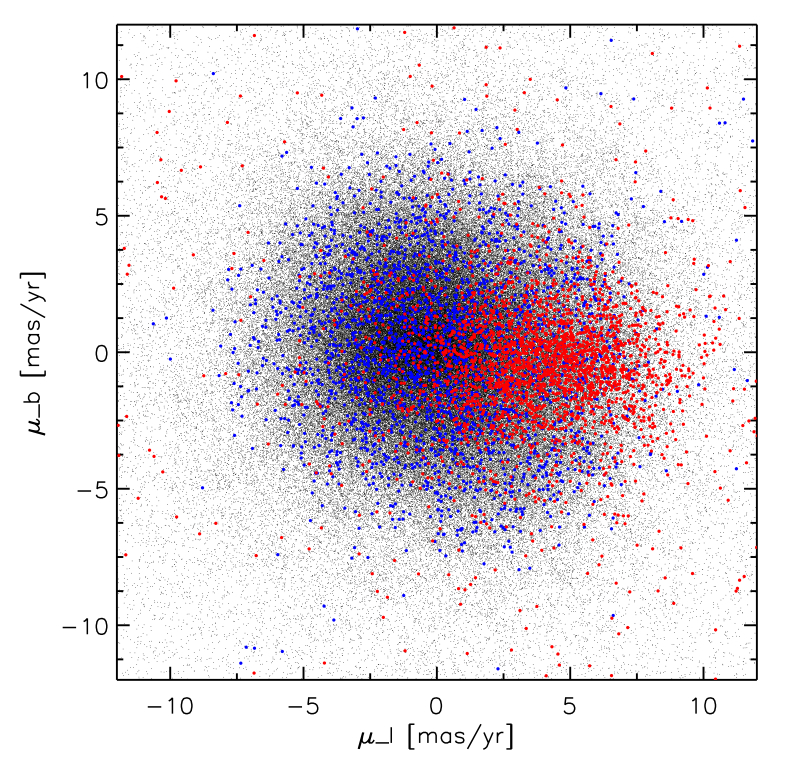}
       \caption{\label{fig2} Proper motion diagram for $\approx$ 200,000 stars down to $F606W \approx$ 28 mag.
       Candidate bulge and disk stars are marked with blue and red dots, respectively (see text for more details).}
     \end{minipage}\hfill
     \begin{minipage}[r]{0.48\textwidth}
      \centering
       \includegraphics[height=6.5cm,width=7cm]{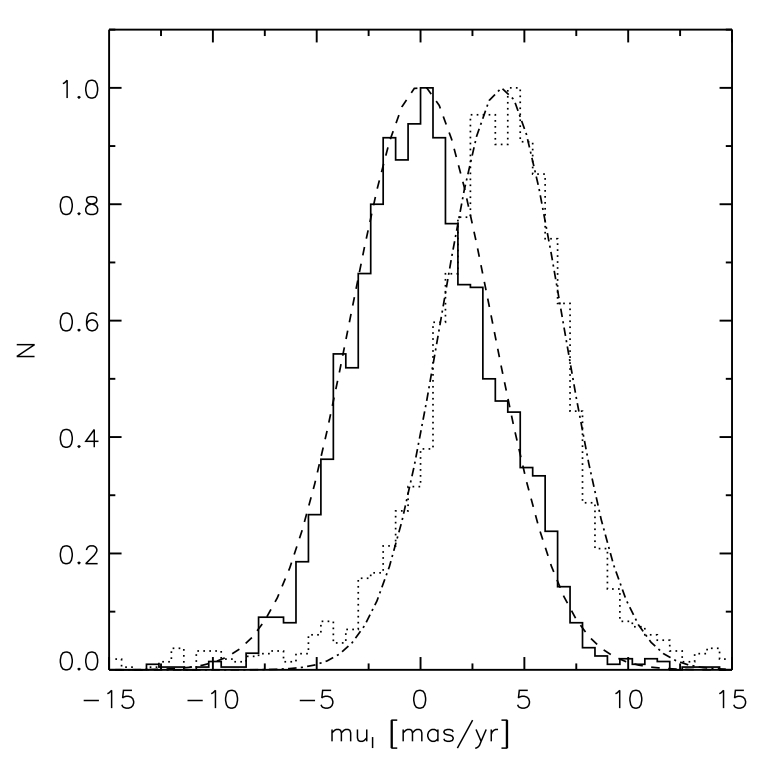}
       \caption{\label{fig3} Proper motions histograms in the Galactic longitude direction for the candidate
       bulge and disk populations. The two Gaussians that fit the distributions are
       over-plotted as dashed lines.}
     \end{minipage}
   \end{figure}

\section{Proper motions and a clean white dwarf bulge sample}\label{pm}
To obtain a clean sample of bulge stars we estimate proper motions (PM) by adopting the 
photometry of 2004 for the SWEEPS field. By comparing the positions of stars in the two epochs
we estimate PMs for a sample of $\approx$ 200,000 stars down to $F606W \approx$ 28 mag, 
with an accuracy $\lesssim$ 0.5 mas/yr in both coordinates. PMs are then projected along the
Galactic coordinates as shown in Fig.~2. From the CMD we select two samples of candidate 
stars belonging to the bulge (blue dots in Fig.~1 and Fig.~2) and to the disk population (red), respectively,
following a procedure similar to that adopted by \citet{clarkson08}.
The PM histograms of the two samples along the longitude ($l$) direction is shown in Fig.~3.
The two populations are clearly separated in this direction, with the bulge peak at $\mu_l \approx$ 0 mas/yr.
We adopted the a cut at $\mu_l \le$-2 mas/yr to select bulge stars as in \citet{clarkson08}: this selection
allows us to keep $\approx$ 30\% of bulge members while the contamination of disk stars is $\approx$ 2\%.
Fig.~4 shows a selected clean bulge CMD: the photometry of bright stars ($F606W < $ 19.8 mag, 
see the dashed line in Fig.~4) is entirely from the 2004 data set, while
fainter stars are from the 2012 data set.
Most of the disk stars have now disappeared but there is still a group of stars
brighter than the MS turn-off (TO) at 17.5 $\lesssim F606W \lesssim$ 19.5 mag and 
0.8 $\lesssim (F606W - F814W) \lesssim$ 1.3 mag. These stars are blue-straggler candidates and
were previously identified in the SWEEPS field by \citet{clarkson11}.
All the EHBs survived the selection and we ended up with a sample of 170 candidate 
bulge WDs. 
\begin{figure}
\center
\includegraphics[width=12truecm,height=11truecm]{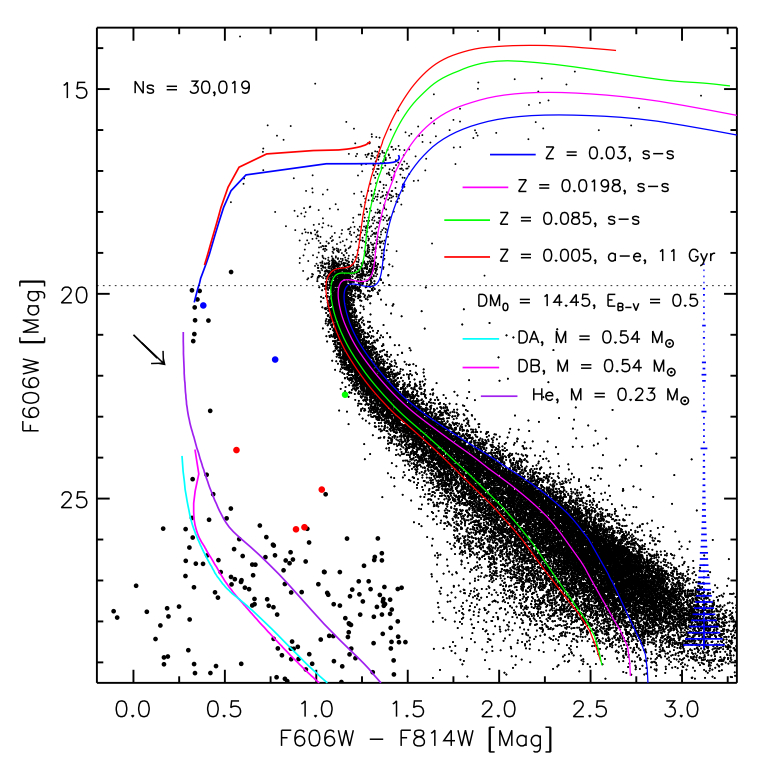} 
\caption{\label{fig4} $F606W,\ F606W - F814W$ CMD of PM selected stars in the SWEEPS field. 
The EHB and the WD stars are marked with larger filled dots.
The solid lines display cluster isochrones and ZAHBs for the same age, $t$ = 11 Gyr, but
different chemical composition (see labeled values).
The adopted true distance modulus and reddening are labeled.
Cooling sequences for CO- and He-core WDs are plotted for different masses.
The arrow marks the reddening vector direction.}
\end{figure}

We validated our photometry over the entire magnitude range (15 $\lesssim F606W \lesssim$ 29 mag)
by adopting the BASTI \citep{pietrinferni06} and Dartmouth \citep{dotter08} stellar-evolution databases.
The solid lines in Fig.~4 show four Dartmouth isochrones for the same age, t = 11 Gyr, and different chemical 
compositions, namely Z = 0.03, Y = 0.288 (blue solid line), Z = 0.0198, Y = 0.273 (purple), Z = 0.0085, Y = 0.256 
(green) and a scaled-solar mixture, and Z = 0.005, Y = 0.252 (red) and an $\alpha$-enhanced mixture ($[\alpha/Fe] = 0.4$).
Zero Age Horizontal Branches (ZAHBs) from the BASTI database are also plotted for the 
most metal-poor (Z = 0.005, red) and the most metal-rich (Z = 0.03, blue) chemical compositions.
We adopted BASTI cooling tracks for DA and DB CO-core WDs with mass M = 0.54 $M_{\odot}$ 
\citep{salaris10} and for He-core WDs we used the models of \citet{althaus09} for a mass of 0.23 $M_{\odot}$.
We adopt a true distance modulus of $\mu = 14.45$ mag and a mean reddening of $E(B-V) = 0.5$ mag 
for the SWEEPS field  \citep{sahu06}. The extinction coefficients for the WFC filters 
are estimated by applying the \citet{cardelli89} reddening relations and by adopting a standard reddening 
law, $R_V = A_V/E(B-V) = 3.1$, finding $A_{F606W} = 0.922 \times A_V$,  $A_{F814W} = 0.55 \times A_V$, and 
$E(F606W - F814W)= 1.14 \times E(B-V)$.

The comparison between theory and observation shows that old (t = 11 Gyr) scaled-solar and $\alpha$-enhanced
isochrones spanning $\sim$ 1 dex in metallicity, -0.6 $\lesssim [M/H] \lesssim$ 0.3, nicely bracket the color range
of the bulge MS and red-giant branch. The residual color spread is probably due to photometric errors, 
differential reddening and the presence of binaries. A more detailed discussion of the bulge CMD will be given in a 
forthcoming paper (Calamida et al. in preparation).

Fig.~4 also shows that cooling tracks for canonical DA (cyan line) and DB (purple) CO WDs are not able to reproduce 
the entire color range of the observed WD cooling sequence in the bulge. An increase in the mean mass of the WDs would move the models towards bluer colors, further increasing the discrepancy.
We then performed several artificial star (AS) tests to estimate the color dispersion due to the photometric errors and to the
reduction techniques adopted. We randomly added to all the images $\approx$ 6,000 artificial WDs, with 
magnitudes and colors estimated by adopting a DA cooling track for CO-core WDs with M = 0.54 $M_{\odot}$.
Images were reduced by following the same procedures and we estimate a completeness of $\sim$ 50\% at $F606W \sim$ 26 mag, and $\sim$ 10\% at $F606W \sim$ 28 mag.
The completeness is strongly dependent on the star color, steeply declining to 0 at $F606W - F814W \sim$ 1.3 mag.
Fig.~5 shows the candidate bulge WDs with a 0.54 $M_{\odot}$ CO-core DA cooling track (24 $< F606W <$ 29.5 mag) 
and a 0.23 $M_{\odot}$  He-core model (23 $< F606W <$ 28 mag) ) over-plotted with the AS color dispersion 
shown for bins of 0.1 mag (black horizontal lines). The red and the green horizontal lines show the same color 
dispersion added to the CO and the He cooling tracks, respectively, but shifted towards faint and redder magnitudes by 
adding 0.15 mag of reddening to each magnitude bin. The red and green sequences then
simulate the presence of WDs affected by differential reddening by a maximum amount of 0.15 mag.
Moreover, the blue line shows the direction of the reddening vector and the two diamonds mark the position of a WD 
in the CMD if affected by 0.5 or 1 mag of extra extinction compared to the other stars. 
This plot shows that differential reddening (added to photometric errors) cannot account for the entire color 
spread of the observed WD cooling sequence in the bulge, unless a large fraction (40\%) of WDs are 
affected by an extra extinction up to 1 mag compared to the other stars. 
Further evidence that this hypothesis is unreasonable comes from the fact that we do not see any evidence 
for extreme differential reddening along the MS (see Fig.~4). 

The comparison with theory and the results of the AS tests suggest that the WD sequence 
color dispersion can be accounted for by assuming that a fraction of WDs in the bulge have lower masses.
These stars are probably He-core WDs, but a fraction of them could also be low-mass CO-core WDs.
The recent theoretical calculations of  \citet{pradamoroni09} showed that CO-core WDs with masses 
down to $\sim$ 0.33 $M_{\odot}$ can form in very high-density environments, when a strong episode of mass-loss 
occurs along the RGB. On the other hand, He-core WDs, in a Hubble time, can only be produced by stars experiencing 
extreme mass loss events such as in compact binaries. This hypothesis is supported by the finding in the SWEEPS
field of a candidate dwarf nova in outburst (green filled dot in Fig.~4), four candidate cataclysmic variables
in quiescence (red dots) , and two ellipsoidal variables (blue dots). 
The details about the identification of these objects and their light curves will be presented in a forthcoming paper 
(Calamida et al., in preparation).

\begin{figure}
\center
\includegraphics[width=12truecm,height=11truecm]{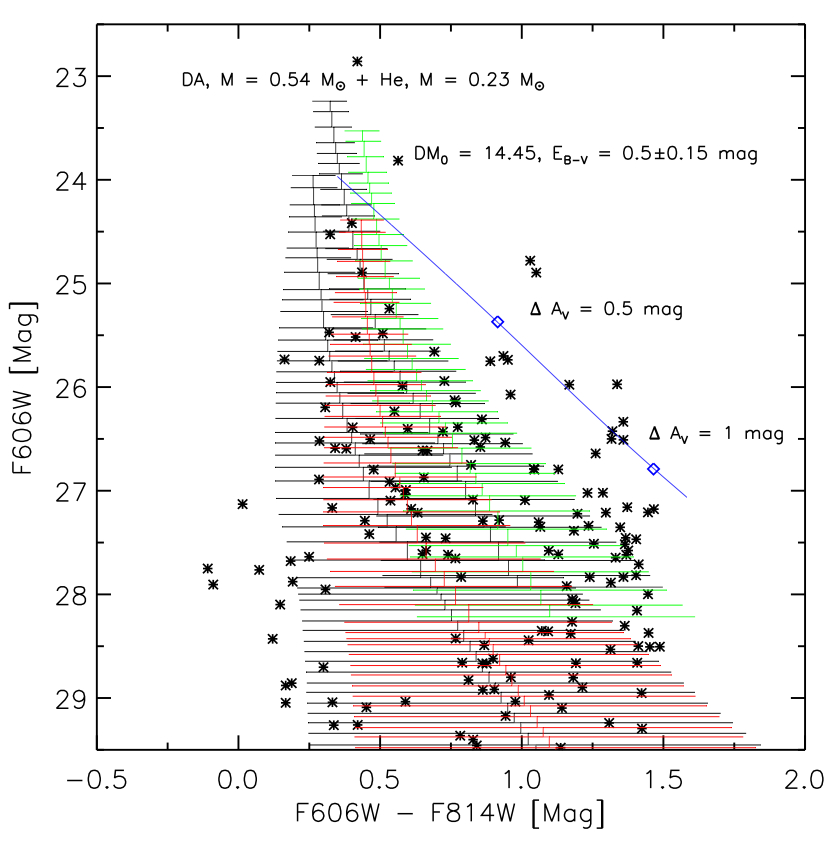} 
\caption{\label{fig5} $F606W,\ F606W - F814W$ CMD of bulge candidate WDs. 
A CO-core DA (black and red lines) and an He-core (black and green) cooling track are plotted with the color dispersion
given by photometric and reduction technique errors and differential reddening added (see text for more details).
The adopted true distance modulus and reddening are labeled.}
\end{figure}

\acknowledgements
Support for this HST Program was provided by NASA through a grant from the Space Telescope Science Institute
(GO-12586), which is operated by the Association of Universities for Research in Astronomy, Incorporated, under NASA contract NAS5-26555.
A. Calamida would like to thank R. J. Foley and A. Rest for their useful suggestions and the interesting discussion about Type Ia SN2002cx.

\bibliography{calamida}

\section{Questions}
\begin{itemize}

\item 1 - For the candidate black hole - subdwarf B system (ellipsoidal variable), have you checked the archival X-ray and
radio data? (Chris Britt)

Yes, unfortunately there was nothing. After getting radial velocities to confirm the nature of the candidate binary system we will
apply for time with Chandra.

\item 2 - You are finding an enhanced population of He-core WD in the Galactic center. There is a type of unusual Type Ia SN 
called SN2002cx which have abundances that suggest He-core WD progenitors.
Would you say that your results of an enhanced population of He-core WD in the center of the Galaxy would predict an 
enhancement of these types of SNe in the centers of other galaxies? (Nick Suntzeff)

Type Ia SN2002cx have a CO-core WD progenitor and a He-rich companion. These SNe have been observed 
only in spiral galaxies up to now, and not in ellipticals (25 vs 0), so the companion is probably a relative young 
Helium burning star \citep{foley10}.
My finding of an excess of He-core WDs in the Galactic bulge would then not influence the result about 
SNe 2002cx. 

\end{itemize}

\end{document}